# Recovering lossless propagation of polaritons with synthesized complex frequency excitation

Fuxin Guan[1,†], Xiangdong Guo[1,2,†], Shu Zhang[2,†], Kebo Zeng[1], Yue Hu[1], Chenchen Wu[2], Shaobo Zhou[1], Yuanjiang Xiang[3], Xiaoxia Yang[2], Qing Dai[2,*], and Shuang Zhang[1,4,*]

[1]New Cornerstone Science Laboratory, Department of Physics, University of Hong Kong, Hong Kong, China

[2]CAS Key Laboratory of Nanophotonic Materials and Devices, CAS Key Laboratory of Standardization and Measurement for Nanotechnology, CAS Center for Excellence in Nanoscience, National Center for Nanoscience and Technology, Beijing 100190, China.

[3]School of Physics and Electronics, Hunan University, Changsha 410082, China

[4]Department of Electrical & Electronic Engineering, University of Hong Kong, Hong Kong, China

[†] These authors contributed equally to this work.

[*]Corresponding author: daiq@nanoctr.cn, shuzhang@hku.hk

## Abstract

Surface plasmon polaritons and phonon polaritons offer a means of surpassing the diffraction limit of conventional optics and facilitate efficient energy storage, local field enhancement, high sensitivities, benefitting from their subwavelength confinement of light. Unfortunately, losses severely limit the propagation decay length, thus restricting the practical use of polaritons. While optimizing the fabrication technique can help circumvent the scattering loss of imperfect structures, the intrinsic absorption channel leading to heat production cannot be eliminated. Here, we utilize synthetic optical excitation of complex frequency with virtual gain, synthesized by combining the measurements taken at multiple real frequencies, to restore the lossless propagations of phonon polaritons with significantly reduced intrinsic losses. The concept of synthetic complex frequency excitation represents a viable solution to compensate for loss and would benefit applications including photonic circuits, waveguiding and plasmonic/phononic structured illumination microscopy.

# Introduction

Polaritons, including surface plasmon polaritons (SPPs)[1–4] and phonon polaritons (PhPs)[5–7], have emerged as a highly promising candidate for constructing nanophotonic circuits[8–14], enabling the development of ultra-compact and high-speed optical devices. Utilizing polaritons in nanophotonics provides a pathway to overcoming the diffraction limit of light, allowing for the manipulation of light at the nanoscale[15–18]. However, the intrinsic losses have hindered many loss-sensitive applications based on polaritons[19–21], including waveguiding[22], bio-sensing[23–26], sub-diffractional limit imaging[27–30], and plasmonic structured illumination microscopy[31,32]. Intrinsic loss negatively impacts polaritons in two primary ways: (I) the propagation distances are significantly reduced, and (II) the dispersion curves of polaritons far from the light cone are significantly blurred, which severely affects the applications of subwavelength behaviors of polaritons.

The intrinsic loss is caused by the imaginary part of dielectric in a material, such as the Ohmic loss in plasmon systems and the lattice vibration relaxation process in phonon-polariton systems[20,33,34]. The most commonly used method to offset loss is to incorporate an external gain medium[35–38]. However, it is very challenging to completely offset the plasmonic loss with gain, and gain compensation[39–41] is susceptible to noises and instabilities.

Complex frequency waves with virtual gain have been proposed to counteract losses in plasmonic/phononic materials for various applications, including super-resolution imaging[42–44], long-range surface plasmon propagation[45], slow light[46–49], coherent virtual absorption[50], light super-scattering[51] and virtual PT symmetry[52]. The concept of complex frequency waves has also been extended to acoustic systems[53]. Complex frequency waves feature temporal attenuation, which requires a precise exponential decay profile in time and time-gated measurements that are challenging for experimental implementation in optics. A recent research has introduced a multi-frequency approach to synthesizing the system response under complex frequency excitation in experiment by combining multiple real frequency measurements[43], which has shown promise in restoring the imaging performance of superlenses that is typically degraded by intrinsic losses.

Here, we demonstrate the recovering of nearly lossless propagation of highly-confined phonon polaritons via the complex frequency approach. The key to recovering lossless propagation lies in the compensation of the imaginary part of in-plane wavevector instead of the imaginary part of permittivity, which is implemented in the mid-infrared regime. Furthermore, our results provide insight into the wave packet dynamics of spatiotemporal evolution of nearly lossless propagation, with the propagation length of complex frequency signal much longer than that of the real frequency. This opens up new possibilities for various applications such as photonic integrated circuits, bio-sensing and microscopy.

**Results and discussion**

Typical materials that support plasmon/phonon polaritons can be described by Drude model, Lorentz model, or multi-Lorentz model[54], which contain loss terms that result in imaginary part of the in-plane wavevector. The wavevector is dependent on frequency, and thus the imaginary part of the complex frequency required for the in-plane wavevector to be real is dependent on its real part. For simplicity, we start with an example of SPPs at an interface, whose the mathematical solution is given by $k = \omega\sqrt{\varepsilon_m}/(c\sqrt{1+\varepsilon_m})$, where we assume the permittivity of plasmonic metal is described by a Drude model $\varepsilon_m = \varepsilon_r - \omega_p^2/(\omega^2 + i\omega\gamma)$. The dissipation term in the denominator introduces decay to the propagation of SPPs. It is intuitive to think that the lossless propagation condition of SPPs could be achieved by fully compensating the loss in the Drude model with a specific complex frequency $\widetilde{\omega} = \omega - i\gamma/2$. However, this is not entirely accurate, as there is still a frequency-dependent term $\widetilde{\omega}$ in the dispersion formula that can result in an imaginary component on wavevectors. The true condition for lossless propagation is indeed given by $Im(k)=0$, which leads to

$$\beta \approx \gamma\omega_p^2\omega^2/(2(\omega_p^2 - \varepsilon_r\omega^2)^2 + 2\varepsilon_r\omega^4) \tag{1}$$

where $\widetilde{\omega} = \omega - i\beta$, $\omega$ corresponds to central frequency and $\beta$ represents the virtual gain, with the details of derivation provided in Supplementary Section 1. The lossless propagation condition for a Lorentz model $\varepsilon_m = \varepsilon_r - \omega_p^2/(\omega^2 + i\omega\gamma - \omega_0^2)$ is also provided in the Supplementary Section 1. This condition ensures that the imaginary part of the wavevector is zero, which in turn implies that there is no spatial growth or decay of the SPPs along the interface. Therefore, to achieve a lossless propagation, it is necessary to find the appropriate values of the imaginary part of frequency.

As an example, we assume a realistic plasmonic metal described by Drude model with permittivity given by $\varepsilon = 5 - \omega_p^2/(\omega^2 + i\omega\gamma)$, where $\omega_p = 1.442 \times 10^{16} Hz$ and $\gamma = 3 \times 10^{14} Hz$. The metal supports SPPs below the plasmon frequency. An infinite long antenna placed on the flat plasmonic metal serves as a SPP source when illuminated by light. The excited field distributions of SPPs are displayed in Fig. 1a. As the frequency increases, the propagation distance of the SPPs decreases, which means that SPPs at higher frequencies have shorter propagation lengths due to their stronger confinement to the interface. The corresponding Fourier distributions of the SPPs are depicted in Fig. 1b. It is shown that the modes at high frequencies are blurred and eventually become invisible. The complex frequency can then be utilized to counterbalance the plasmonic loss for achieving lossless propagation. The imaginary frequency $f_i$ for lossless propagation condition is calculated by using Eqn. 1 and plotted in Fig. 1c. It is shown that as the real frequency increases, the imaginary frequency also increases monotonically to overcome the propagation loss. However, a complex frequency wave (temporally attenuated wave) is unphysical as the energy approaches infinity as time approaches negative infinity. Therefore, a truncation at the start of time is necessary to rationalize the complex frequency wave. The truncated complex frequency wave with temporal attenuation $e^{-i\widetilde{\omega}t}\theta(t)$ can be transformed into real frequency domain via Fourier transformation, and the corresponding spectrum distribution has a Lorentzian lineshape $1/(i\widetilde{\omega} - i\omega')$. Thus, the complex frequency field distribution can be synthesized via the linear combination of the real frequency field distributions as[43],

$$E(\widetilde{\omega}, r, t) = \sum_i E(\omega'_i, r) e^{-i\omega'_i t} \Delta\omega / (2\pi i(\widetilde{\omega} - \omega'_i)) \qquad (2)$$

where the imaginary part of the complex frequency is obtained from Eq. 1. The dispersion under complex frequency excitation is shown in Fig. 1d, which fully recovers the dispersion at higher frequencies. The corresponding mathematical derivation and temporal evolution information are shown in Supplementary Section 2 and Fig. S1. The dynamic evolution of the complex frequency excitation at $\tilde{f} = (700 - 6.73i)$ THz (its real part is indicated by the bottom dashed line in Fig. 1d) at different moments is depicted as Fig. 1e. An interesting observation is that as time increases, the propagation distance extends linearly, but the amplitude remains uniform at different positions. This provides a direct visualization of lossless propagation of surface plasmon polaritons at a complex frequency, which is in sharp contrast with that at the real frequency. The corresponding spatial field distribution at $t = 5.5 \times 10^{-2}$ ps is illustrated in the upper panel in Fig. 1f. We also plot the field distribution at a higher complex frequency $\tilde{f} = (850 - 17i)$ THz (indicated by the upper dashed line in Fig. 1c), with the result shown in the lower panel in Fig. 1f, which also shows much slower decay in comparison to the real frequency case.

The compensation of loss applies not only to metals described by Drude model, but also to materials with more complicated dielectric functions, such as van der Waals materials that support PhPs. There is always a specific complex frequency solution for a PhP that satisfies $Im(k)=0$. Here we consider hexagonal boron nitride (hBN) film, which supports in-plane isotropic PhPs[54–58], and show that complex frequency can be utilized to compensate its intrinsic loss for observing lossless propagation of the phonon polariton. The experimental setup, based on the s-SNOM technique, is illustrated in Fig. 1g. A long gold antenna placed on the hBN film is used to launch the 1D PhPs (details in Supplementary Fig. S2). The electric field distributions are measured at frequencies ranging from 1421 cm$^{-1}$ to 1503 cm$^{-1}$, with a step of 2 cm$^{-1}$, with all field plots provided in Supplementary Fig. S3. Two field distributions with real frequencies of 1451 cm$^{-1}$ and 1477 cm$^{-1}$ are chosen as the central frequencies to synthesize complex frequency responses at $\tilde{f}_1 = (1451 - 4.5i)$ cm$^{-1}$ and $\tilde{f}_2 = (1477 - 6i)$ cm$^{-1}$. The corresponding imaginary part represents the optimized value for each frequency. The field distributions of the two real frequencies and the two complex frequencies are displayed in Fig. 1h and 1i, respectively. The experimental results demonstrate that while the propagation at the real frequencies experiences strong attenuation, the propagation of polariton at the complex frequencies is nearly non-dissipative.

We next apply the complex frequency method to investigate the temporal evolution of more complicated field distributions supported by a thin film of van der Waals crystal $\alpha - MoO_3$ ($MoO_3$), which is highly anisotropic and supports natural in-plane hyperbolic polaritons[59–62]. A gold antenna is placed on the $MoO_3$ film to excite the PhPs, with the atomic force microscope (AFM) image displayed in Fig. 2 (details in Supplementary Fig. S4). The near-field distributions on the $MoO_3$ interface measured by SNOM at different frequencies are shown in Fig. 2. The overall frequency range spans from 891 to 943 $cm^{-1}$ with an interval of 1 $cm^{-1}$. Only nine field plots are presented in the figure while the others are displayed in Supplementary Figs. S5 and S6. The field distribution variation exhibits a characteristic hyperbolic propagation behavior with a concave wavefront. With the increase of the frequency, the wavelength decreases with a stronger field confinement, and meanwhile the propagation becomes more attenuated. At all measured frequencies, the decay lengths of the polaritons are less than two wavelengths due to the significant intrinsic loss of the material.

The synthesized electric field distributions at the complex frequency $(910 - 6.5i)$ $cm^{-1}$ and at different temporal moments are depicted in Fig. 3a, with a temporal step of 0.53 ps. Interestingly, for small *t*, the noise is largely smoothed by the summation of multi-frequency signals. As time goes on, the wave travels farther and farther up, extending significantly beyond the decay length of two wavelengths at real frequencies. The 2D field plot reaches its optimum at around 4.24 ps with the propagation length of 8.7 μm (along vertical direction), while the corresponding propagation length of the central real frequency is only 1.6 μm. Thus, the propagation length under such complex frequency approach is increased by more than four times compared with that of the central real frequency. Other complex frequency results are displayed in the supplementary Fig. S7, which all show significantly longer propagation distance than the corresponding real frequencies. By further increasing *t*, the noise starts to become dominant, as the signal continues to attenuate in the time domain. As time exceeds 4.77 ps, the field distribution begins to exhibit chaotic behavior. The dynamic movie of spatiotemporal evolution is shown in Supplementary Video 1. Thus, there exists a trade-off between a longer propagation and the signal noise ratio when deciding the optimal time for constructing waves in complex frequency domain.

The effect of the number of frequency points is also investigated as shown in Fig. 3b, with the frequency step fixed at 1 cm$^{-1}$ and the temporal snapshot fixed at 4.24 ps. It is evident that the field plot with 9 frequency points is already significantly improved compared to that of real frequency. As the number of frequencies increases to 21, the decay length increases significantly. However, further increasing the number to 39 does not provide obvious improvement.

We further apply the complex frequency approach to investigate the interference behavior of PhPs. Two circular antennas with different diameters (~0.8 μm and ~3 μm) are fabricated on the $MoO_3$ film (details in Supplementary Fig. S8) to excite the phonon polaritons, as shown in Fig. 4a. The field distribution inside the white dashed box is scanned using an s-SNOM probe, with the amplitude and real part of the electric field distribution at a frequency of $f = 990$ cm$^{-1}$ shown in the top and bottom panel of Fig. 4b, respectively. The PhPs emanating from the two antennas propagate towards each other, but they do not form discernible interference fringe due to the strong attenuation. By synthesizing the field plots at the complex frequency $\tilde{f} = (990 - 2i)$ cm$^{-1}$, the intrinsic loss in the $MoO_3$ film is compensated for, thereby enabling longer propagation of PhPs. This facilitates the formation of clear interference patterns between the antennas, as shown by Fig. 4c. The dynamic behavior is displayed in the Supplementary Video 2, which exhibits the interesting features of a negative phase velocity and a slow group velocity, because of the negative slope of dispersion of the PhPs across this frequency range. All the measured field distributions at different real frequencies for constructing the synthesized complex frequency field plot are provided in the Supplementary Fig. S9 and S10.

**Conclusions**

In conclusion, we have implemented the complex frequency approach to significantly enhance the decay length of highly-confined phonon polaritons, restoring nearly lossless propagation that is only limited by the noise level. This concept is general and can be easily extended to other frequency regimes and to other types of waves, including acoustic and elastic waves. The ability to achieve nearly lossless propagation of phonon polaritons has significant implications for a wide range of applications. In particular, the lossless propagation of phonon polaritons could be leveraged in the development of photonic integrated circuits, where the ability to transmit signals over longer distances is critical for ensuring the performance and reliability of the circuit.

## Methods

**Nanofabrication of the devices.**

Mechanically exfoliated high-quality hBN and α-MoO3 flakes were obtained from bulk crystals synthesized, and were subsequently transferred onto either commercial 280 nm SiO2/500 μm Si wafers (SVM) or gold substrates using a deterministic dry transfer process with a polydimethylsiloxane (PDMS) stamp. Gold antennas were patterned onto specific MoO3 flakes and hBN flakes by coating approximately 350 nm poly (methyl methacrylate) (PMMA) 950K lithography resist and utilizing 100 kV electron-beam lithography (Vistec 5000+ES). Afterward, 10 nm Cr and 100nm Au were evaporated and deposited using electron-beam in a vacuum chamber at a pressure of less than $1\times10^{-6}$ torr. To remove the PMMA from the samples, a hot acetone bath in water (60 °C, 40 min) was utilized, followed by a gentle washing with isopropyl alcohol (IPA) for 3 min, and drying with flowing nitrogen gas. The morphologies of the samples are illustrated in Figs. S2, S4 and S8 of the Supplementary Information.

**Scanning near-field optical microscopy measurements.**

A scattering scanning near-field optical microscope (Neaspec) equipped with a widely tunable mid-IR OPO laser (550-2200 cm-1) was utilized to visualize near optical fields. The probes, initially designed for metalized atomic force microscope (AFM) with an apex radius of around 10-20 nm (Nanoworld), were employed, and the tip-tapping frequency and amplitudes were set to roughly 270 kHz and 50-100 nm, respectively. By combining interferometric measurement of the light backscattered by the tip with demodulation of the detector signal at multiples nΩ (usually second or third harmonics), we achieved background-free recording of the near-field scattering. Phase modulation of the reference beam allowed for the separation of the signal amplitude |En| and phase ϕn. The p-polarized IR light was focused through a parabolic mirror onto both the tip and sample at an angle of approximately 52° relative to the tip axis.

**Numerical simulation of excitation with SPPs.**

The electromagnetic simulations performed using a commercial full-wave finite element methods software called Comsol Multiphysics. A flat interface of air/dispersive Drude model with $\varepsilon_m = 5 - (1.44e4 \text{ THz})^2/(\omega^2 + (3e2 \text{ THz})i\omega)$ supports SPPs composed of both *x* and *y* polarized electric field. A metal antenna, 80 nm width, was placed on the 2D interface, and an *x*-polarized plane wave was normally illuminated onto the antenna to excite the SPPs on the interface. Perfectly matched layer boundaries were employed in both the *x* and *y* direction in the 2D simulation to absorb the clutter waves. The simulation frequency range was swept from 500 to 1030 THz with the frequency gap of 0.5 THz. We recorded the *y*-polarized electric field of each frequency just atop the interface to synthesize the complex frequency wave.

## Acknowledgements

This work was supported by New Cornerstone Science Foundation and the Research Grants Council of Hong Kong AoE/P-701/20, 17309021 (to Shuang Zhang); National Key Research and Development Program of China grant 2021YFA1201500 (to Q.D.); National Natural Science Foundation of China (U2032206 and 51925203 to Q.D.; 52022025 to X.Y. and 52102160 to X.G.).

## Author contributions

Shuang Zhang, F.G., Q.D. and X.G. conceived the project. F.G. and X.G. designed the experiments. X.G., and Shu Zhang prepared the experimental samples and performed the s-SNOM experiments with the help of C.W.; F.G. and X.G. solved the experimental data and performed the simulation. F.G., X.G., Shu Zhang, K.Z., Y.H., Y.X., Shaobo Zhou., X.Y., Q.D. and Shuang Zhang participated in the analysis of the results. F.G., X.G. and Shuang Zhang wrote the manuscript with input from all authors. All authors contributed to the discussion.

## Competing interests

The authors declare no competing interests.

## Data availability

The data that support the findings of this study are available within the paper and the Supplementary Materials.

## Code availability

The code that supports the findings of this study is available from the corresponding author upon reasonable request.

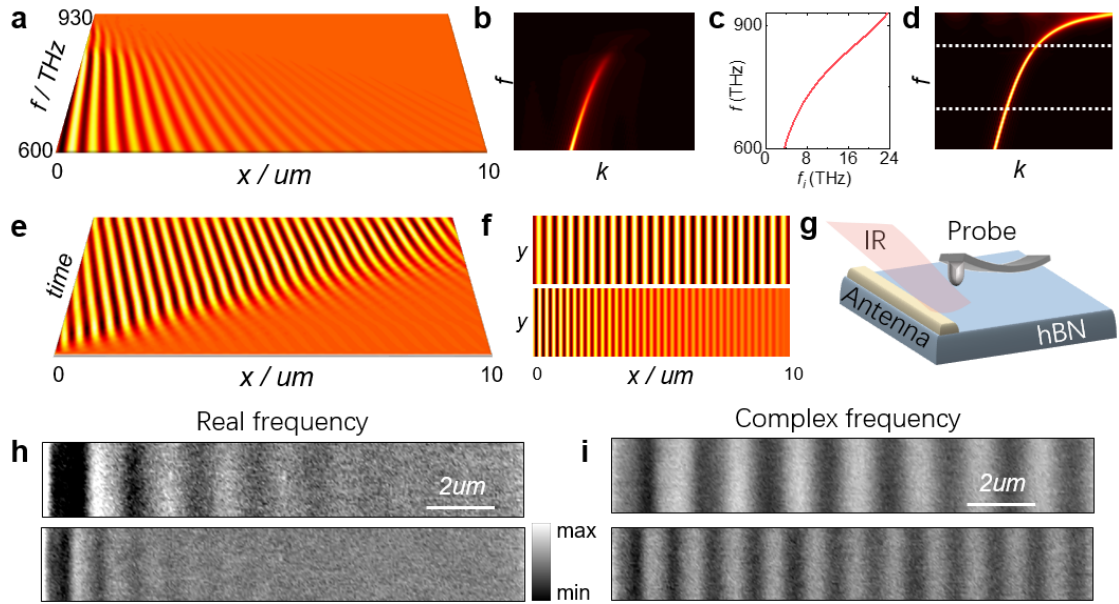

**Fig. 1. Concept of lossless propagation by synthetic wave of complex frequency. a,** SPPs propagation distribution as a function of frequency scattered by an infinitely long antenna. **b,** The corresponding Fourier distribution of the propagation field. **c,** The imaginary frequency as function of real frequency derived by Eq. 1. **d,** The synthetic Fourier distribution in complex frequency domain under $Im(k)=0$ condition. **e,** The spatiotemporal dynamics of complex frequency of the real frequency are given by the lower dashed line position in (c). **f,** The electric field distribution at the two dashed line frequencies in (c). **g,** Schematic of s-SNOM experimental setup with a hBN flake laid on a gold antenna. Illumination on the antenna excites the PhPs and electric field distribution is measured by the probe. **h**, Two real part near-field distributions at the real frequencies of 1451 cm$^{-1}$ and 1477 cm$^{-1}$, respectively. **i,** The corresponding complex frequency near-field distributions. Bright and dark colors correspond to maximum and minimum values.

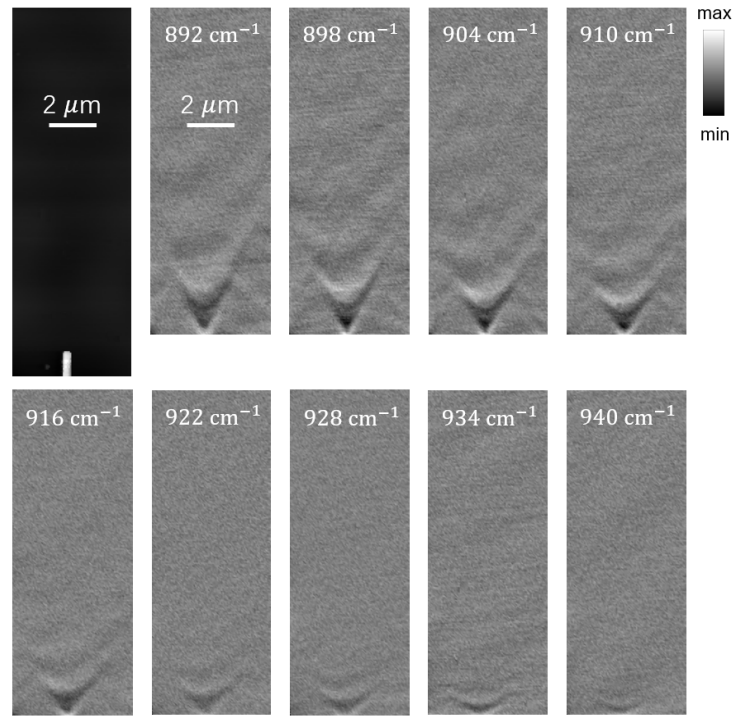

**Fig. 2. AFM and s-SNOM imaging in the vicinity of an antenna on an $MoO_3$ film that supports hyperbolic PhPs.** The depicted real part near-field distributions are labeled with their corresponding real frequency in the inset. Bright and dark colors correspond to maximum and minimum values.

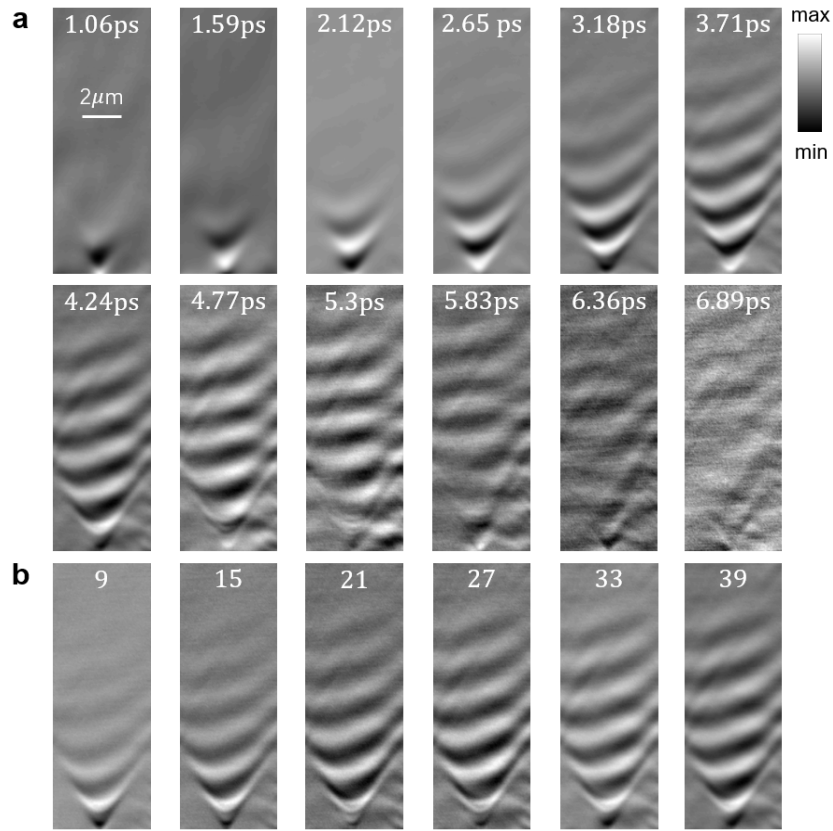

**Fig. 3. The synthetic s-SNOM imaging with a complex frequency of $\tilde{f} = (910-6.5i)$ cm$^{-1}$ in different transient snapshots and different number of synthesized real frequencies.** a, The synthetic field in different transient with a total frequency number of 53, ranging from 891 to 943 cm$^{-1}$. The temporal interval between the photos is 0.53 ps, and the starting time is 1.06 ps. b, The imaging patterns with different synthetic frequency numbers, where the central frequency is fixed at 910 cm$^{-1}$ and the frequency interval is fixed at 1 cm$^{-1}$. The corresponding frequency numbers are labelled in the inset.

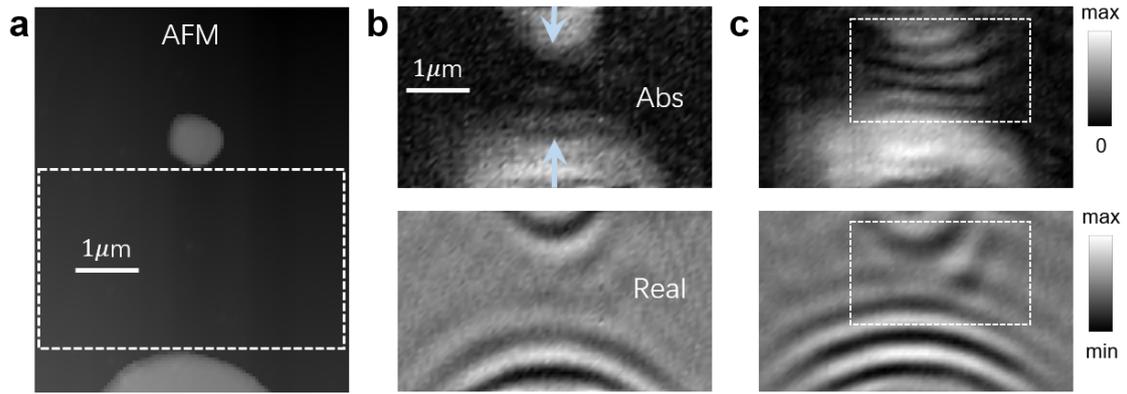

**Fig. 4. Investigation of interference of PhPs from two different circular antennas on MoO$_3$ flake.** a, The AFM image of the experimental sample. The bulges corresponds to the position of the antenna. b, The amplitude and real part of the near-field distribution at 990 cm$^{-1}$ measured by s-SNOM. The scan region is in the dashed box of (a). c, The amplitude and real part of the near field at the synthetic complex frequency ($\tilde{f}$=990-2i) cm$^{-1}$ with a total frequency number of 26, the corresponding temporal snapshot is 11.14 ps. The dashed boxes correspond to the strong interference region.